# Recovery of graded index profile by cubic spline function

**Weijun Liao, Xianfeng Chen*, Yuping Chen, Yuxing Xia and Yingli Chen**

*Department of Physics, The State Key Laboratory on Fiber-Optic Local Area Communication Networks and Advanced Optical Communication Systems, Shanghai JiaoTong University. 800, Dong Chuan Road, Shanghai, 200240, People's Republic of China*

**Abstract:**

We present in this paper a method to recover the refractive index profile of graded waveguide from the effective indices by cubic spline interpolation function. It is proved by numerical analysis of several typical index distributions that the refractive index profile can be reconstructed closely to its exact profile with the presented interpolation model. This method can reliably retrieve index profile of both more-mode (more than 4 guiding mode) and fewer-mode (2-4) waveguides.

## I. INTRODUCTION

The refractive index profile plays a fundamental role in a graded index waveguide as it can give significant information on the waveguide propagation properties, thus the determination of the profile has attracted considerable interests at all times. The techniques used to solve this problem are classified to destructive and non-destructive method. In destructive approaches, samples must be processed before the measurement, such as reflectivity profiling [1], ellipsometry [2]. The non-destructive method takes advantage of intact waveguide sample. The most commonly used method, such as inverse WKB (IWKB) method [3] and improved IWKB method [4], is to reconstruct index profile from measured effective indexes.

The methods are also developed [5] to deal with waveguides with fewer modes. However, WKB analysis is an approximate method, which will unavoidably restrict the accuracy of its inverse methods. In the case of waveguides supporting fewer modes other techniques such as changing the cover layer should be added, making the recovery process very complicated. In this paper, an interpolation method with cubic spline function [6] based on an exact analytic transfer matrix (ATM) method [7] is introduced to recover the refractive index profile from effective indexes. The cubic splines can get a quite smooth profile as it ensures that the first and second derivatives at each interpolation point are continuous, which is consistent with the real situation. With an iterative procedure this method can recover the exact index profile with good accuracy.

## II. RECOVERY PROCESS

A series of mode indexes $n_i\,(i = 0,1,...,k)$ of graded-index waveguide can be measured by m-lines method [8]. A set of increasing coordinate $x_i\,(i = 0,1,...,k)$ is assumed arbitrarily to mode index $\{n_i\}$, and the surface index is $n_c$ ($x_c = 0$), which is unknown and should be determined. Now we have a sets of points [($n_c$, 0), ($n_0, x_0$), ($n_1, x_1$), …, ($n_k, x_k$)]. These points can be interpolated by cubic spline functions with a smooth profile. In order to work out the unknown surface index $n_c$, the point ($n_0, x_0$) is picked out from the sets of points. So the series of points [($n_c$, 0), ($n_1, x_1$), ($n_2, x_2$), …, ($n_k, x_k$)] except point ($n_0, x_0$) are finally used to fit the cubic splines. When $x \geq x_k$, an exponential profile is assumed, thus according to the interpolation theory of cubic spline function, the index profile can be described as follows:

$$n(x)=\begin{cases} M_0\dfrac{(x_1-x)^3}{6h_0}+M_1\dfrac{x^3}{6h_0}+(n_c-\dfrac{M_0h_0^2}{6})\dfrac{x_1-x}{h_0}+(n_1-\dfrac{M_1h_0^2}{6})\dfrac{x}{h_0}, & 0\le x\le x_1 \\ M_i\dfrac{(x_{i+1}-x)^3}{6h_i}+M_{i+1}\dfrac{(x-x_i)^3}{6h_i}+(n_i-\dfrac{M_ih_i^2}{6})\dfrac{x_{i+1}-x}{h_i}+(n_{i+1}-\dfrac{M_{i+1}h_i^2}{6})\dfrac{x-x_i}{h_i}, & \\ & x_i\le x\le x_{i+1}, i=1,2,...,k-1 \\ n_s+b\cdot\exp(-ax), & x\ge x_k \end{cases} \quad (1)$$

Where $h_0=x_1$, $h_i=x_{i+1}-x_i$, $(i=1,2,...,k-1)$, and $M_i(i=0,1,...,k)$ is the second derivative of $n(x)$ at the sets of points $[(n_c, 0), (n_1, x_1), (n_2, x_2), \ldots, (n_k, x_k)]$. $n_s$ is the refractive index of the substrate. Parameters $a$ and $b$ of the exponential profile beyond $x_k$ is determined by two points $[(n_k, x_k),(n_{k+1}, x_{k+1})]$, where $x_{k+1}\gg x_k$ and $n_{k+1}=n_s+\varepsilon$, $\varepsilon$ is very small. It is proved by calculation that $x_{k+1}=5x_k$ to $10x_k$ and $\varepsilon=1\times10^{-10}$ to $1\times10^{-7}$ will have little influence to the final results. Reasonably and technically, we set $x_{k+1}=5x_k$ and $\varepsilon=1\times10^{-7}$. The series of second derivatives $M_i(i=0,1,...,k)$ is determined by the following system of linear equations:

$$\begin{bmatrix} 2 & 1 & & & \\ \mu_1 & 2 & \gamma_1 & & \\ & \ddots & \ddots & \ddots & \\ & & \mu_{k-1} & 2 & \gamma_{k-1} \\ & & & 1 & 2 \end{bmatrix}\begin{pmatrix} M_0 \\ M_1 \\ \vdots \\ M_{k-1} \\ M_k \end{pmatrix}=\begin{pmatrix} d_0 \\ d_1 \\ \vdots \\ d_{k-1} \\ d_k \end{pmatrix} \quad (2)$$

Where $\mu_i=\dfrac{h_{i-1}}{h_{i-1}+h_i}$, $\gamma_i=1-\mu_i$, $(i=1,2,...,k-1)$, $d_0=\dfrac{6}{h_0}[\dfrac{n_1-n_c}{h_0}-n'(0)]$, $n'(0)$ is the first derivative of $n(x)$ at the surface $x=0$. $n'(0)$ will be given and may be changed, which will be discussed in detail in latter part.

$$d_1=6\frac{\dfrac{n_1-n_c}{h_0}-\dfrac{n_2-n_1}{h_1}}{-x_2} \quad (3)$$

$$d_i=6\frac{\dfrac{n_i-n_{i-1}}{h_{i-1}}-\dfrac{n_{i+1}-n_i}{h_i}}{x_{i-1}-x_{i+1}}, \qquad i=2,...,k-1 \quad (4)$$

$$d_k = \frac{6}{h_{k-1}}[n'(x_k) - \frac{n_k - n_{k-1}}{h_{k-1}}], \tag{5}$$

where $n'(x_k)$ is the first derivative of $n(x)$ at $x = x_k$, which is equal to the first derivative of the exponential profile at $x = x_k$ under continuous condition, and $n'(x_k) = -ab \cdot \exp(-ax_k)$.

Because of the surface index $n_c$ is unknown we need another known condition added to the system of linear equations, so that all of the unknown parameters can be solved. This added condition is that the point ($n_0, x_0$) lies in the curve from ($n_c$, 0) to ($n_1, x_1$), which can be depicted as:

$$n_0 = M_0 \frac{(x_1 - x_0)^3}{6h_0} + M_1 \frac{x_0^{\ 3}}{6h_0} + (n_c - \frac{M_0 h_0^{\ 2}}{6}) \frac{x_1 - x_0}{h_0} + (n_1 - \frac{M_1 h_0^{\ 2}}{6}) \frac{x_0}{h_0} \tag{6}$$

Finally, for an arbitrarily given series of $\{x_i\}$, the index distribution can be fitted with a very smooth profile according to the continuous first and second derivatives of all cubic spline functions, even at the interpolation points [($n_1, x_1$), ($n_2, x_2$), …, ($n_{k-1}, x_{k-1}$)]. Then we use an exact analytic transfer matrix (ATM) method to solve the waveguide with this fitting index profile to get the corresponding effective indexes $n_i^{cal}$ ($i = 0,1,...,k$). Here we just give the dispersion equations of ATM method, which is presented as follows:

$$\int_0^{x_t} k(x)dx + \Phi(\Gamma) = m\pi + \tan^{-1}(\frac{p_0}{k_1}) + \tan^{-1}(\frac{p_t}{k_l}) \qquad (m = 0,1,2,...)$$

Where $x_t$ is turning point of the monotonically decreasing index profile under $m$th mode. $k(x) = [k_0^{\ 2} n^2(x) - \beta^2]^{1/2}$, $k_0 = 2\pi / \lambda$, and $\lambda$ is the wavelength in air. $\beta$ is the propagation constant, and $\beta = k_0 n(x_t)$. $p_0 = (\beta^2 - k_0^{\ 2} n_a^{\ 2})^{1/2}$, $n_a$ is the refractive index of the cover layer. $k_1 = [k_0^{\ 2} n^2(0) - \beta^2]^{1/2}$, $n(0)$ is the surface index of the waveguide. $p_t = (\beta^2 - k_0^{\ 2} n_{eq}^{\ \ 2})^{1/2}$, $n_{eq}$ is characterized as equivalent refractive index beyond $x_t$. $k_l \to 0$ as $l \to \infty$, $m$ is the mode number, $\Phi(\Gamma)$ is interpreted as the phase contribution of the subwaves. Details about this ATM method are in [7].

After solving the interpolated index distribution a new series of effective index $\{n_i^{cal}\}$ is obtained. Simultaneously, a new series of $\{x_i\}$ is acquired. We define and calculate the departure of the effective indexes between calculated values and exact values as $\Delta = \sum_{i=0}^{k} (n_i^{cal} - n_i)^2$. We can evaluate this deviation, if it is still large enough, we substitute the new series of $\{x_i\}$ to Equation (1), and a new index profile can be fitted with interpolation on the new series of points. Series effective indexes $\{n_i^{cal}\}$ and $\{x_i\}$ can be determined by the solution of the new index distribution in the waveguide with ATM method. Repeating the above approach the deviation $\Delta$ will get smaller and smaller, that is, this iteration process is convergent and the profile is approaching to the real profile. When $\Delta$ is close to zero, the refractive index profile is finally acquired.

It should be noted that for the facility of the iteration procedure the first series of $\{x_i\}$ should be chosen so that the waveguide of the fitting profile have enough guiding modes. We can choose $\{x_i\}$ as arithmetic series and the interval is chosen to be $5\lambda \sim 8\lambda$, where $\lambda$ is the wavelength in air.

In previous discussion we have noted that $n'(0)$, the first derivative of $n(x)$ at the surface $x = 0$, is given and may be altered. For a given $n'(0)$, the obtained profile may be impractical. Because the second derivative can reflect the concave and convex character of the curve, practicability of the index distribution can be judged by investigation on the series of second derivatives $M_i (i = 0,1,...,k)$, as for a practical refractive index profile the signs of $M_i (i = 0,1,...,k)$ have special rules. We have the judge rule that every sign of $\{M_i\}$ before the first positive sign should be negative and every sign of $\{M_i\}$ after the first positive sign should be positive, or all of signs of $\{M_i\}$ are positive. As we know, for most of graded index profiles such as Gaussian profile, error function profile, Fermi profile and exponential profile, $n^{'}(0)$ is less than or equal to zero. At first, we set $n^{'}(0)=0$, and under this value we can get a convergent profile. Then, if the signs of $\{M_i\}$ in the profile satisfy the judge rule, the practical index profile is acquired, and if the signs of $\{M_i\}$ don't satisfy the judge rule, $n^{'}(0)$ should be decreased until $\{M_i\}$ from the calculated profile accord with the judge rule. It is proved by simulation results that from a series of effective indexes $\{n_i\}$ the index distribution can be precisely recovered very close to its exact profile with the iteration approach and the judge rule.

## III. NUMERICAL RESULTS AND COMPARISON

In order to investigate the reliability of this method, we give some typical examples of graded index profiles such as Gaussian, error function, Fermi and exponential profile. The cover layer is uniformly the air with index $n_a = 1.0$. All numerical simulations are performed with wavelength of 632.8 nm and unit of length in micrometer. Every index profile in waveguides is also calculated under different mode numbers to verify the universality of the current method.

Firstly we consider waveguides with index distribution of Gaussian profile, which is described as $n(x)=1.5+0.025\exp(-x^2/D_{\text{gauss}}{}^2)$, $D_{\text{gauss}}$ varies from $3\mu m$ to $7\mu m$, allowing the waveguides to support 3~7 modes. All waveguides have uniform surface index with the value of 1.525. The numerical results are demonstrated in Fig.1(a). A good superposition can be observed between the exact and calculated profiles in the guiding region. Some discrepancy exists in the retrieved profiles after the last mode because an exponential profile is assumed when $x \geq x_k$ in our calculations. The next example is implemented in waveguides with error function profile, which is depicted as $n(x) = 2.2 + \frac{n_{\text{erf}}}{2}[erf(\frac{1.0+x}{3.6}) + erf(\frac{1.0-x}{3.6})]/erf(\frac{1.0}{3.6})$, where $n_{\text{erf}}$ =0.01, 0.02, 0.035, 0.045, corresponding to surface index, can stimulate 3, 4, 5 and 6 modes, respectively. The results are shown in Fig.1(b). The recovered values of $n_{\text{erf}}$ for each waveguide are 0.00998, 0.02012, 0.03516, 0.04518, which agree well with the exact values. Similar results are also obtained with exponential profile $n(x) = 2.47+0.01\times\exp(-x/D_{\exp})$ and slowly varying Fermi profile $n(x)=1.735+n_{\text{fermi}}/\{1+\exp[(x-2.5)/0.7]\}$, as shown in Fig. 1(c) and Fig.1(d).

To further verify the reliability of our method, we consider waveguides with only two guiding modes. Because of the invalidation of the judge rule on only 2 derivatives of the second order we just predict Gaussian, error function and slowly changing Fermi profiles except exponential profile, which can uniformly be calculated with $n'(0) = 0$ . The effective indexes of $TE_0$ and $TE_1$ of Gaussian profile $n(x) = 1.5 + 0.018\times\exp(-x^2/2.8^2)$ are 1.5108 and 1.5033, respectively. The calculated result is shown in Fig.2 (a), which also agrees well with the exact profile as the above simulations. Fig.2 (b) is achieved from error function profile $n(x) = 1.5+0.018\times 1/2\times\{\text{erf}[(1.0+x)/2.3]+\text{erf}[(1.0-x)/2.3]\}/\text{erf}(1.0/2.3)$ , which supports two guiding modes with effective indexes of 1.5099 and 1.5020. The obtained surface index is 1.51792, very close to the exact value. Although the obtained index distribution of Fermi profile $n(x) = 1.5+0.018/\{1+\exp[(x-2.5)/0.7]\}$ expects a less accurate achievement, the result still give a good prescription on the index amplitude and depth of the waveguide, as shown in Fig.2(c).

In comparison with IWKB method we consider a Gaussian profile $n(x) = 1.5 + 0.025\times\exp(-x^2/5^2)$ supporting 5 guiding modes. The effective indexes are 1.519 95, 1.513 75, 1.508 43, 1.504 11, 1.501 07, respectively. Recovered profiles by IWKB and the current method are shown in Fig.3. It can be found that the current method can estimate the index profile in more accuracy than IWKB method, especially in the recovery of surface index.

## IV. CONCLUSION

In summary, we have demonstrated that the refractive index profile of graded index waveguide can be smoothly recovered in good accuracy with cubic spline interpolation functions based on exact ATM method and simply iterative approaches. This method can uniformly predict multimode waveguides with more modes and fewer modes reliably from the effective indexes. It can also be used to investigate untypable profiles from measurement of effective indexes by changing the value of $n^{'}(0)$ and investigation on the second derivatives. The explicit analysis provides a reliable and convenient technique in the approach of graded index profiling.

Figure Captions:

Fig.1(a) Recovery of Gaussian distribution by cubic spline function. $D_{\mathrm{gauss}}$ varies from $7\mu m$ to $3\mu m$, allowing the waveguides to support 7~3 modes ( from the top down ).

Fig.1(b) Recovery of error function distribution by cubic spline function. $n_{\mathrm{erf}}$ = 0.045, 0.035, 0.02, 0.01 stimulate 6 modes, 5 modes, 4modes, 3 modes respectively (from the top down).

Fig.1(c) Recovery of exponential distribution by cubic spline function. The waveguides with $D_{\exp}$ of $5.5\mu m$, $4.3\mu m$, $3.4\mu m$, $2.5\mu m$ support 6,5,4,3 guiding modes respectively (from the top down).

Fig.1(d) Recovery of slowly varying Fermi distribution by cubic spline function. $n_{\mathrm{fermi}}$ =0.1, 0.07, 0.05, 0.03 correspond with 6,5,4,3 stimulated guiding modes respectively (from the top down).

Fig. 2 Recovery of two-mode waveguide with (a) Gaussian profile, (b) error function profile, (c) slowly varying Fermi profile

Fig.3 Comparison of recovered profiles from effective indexes by IWKB and the current method.

Fig.1(a)

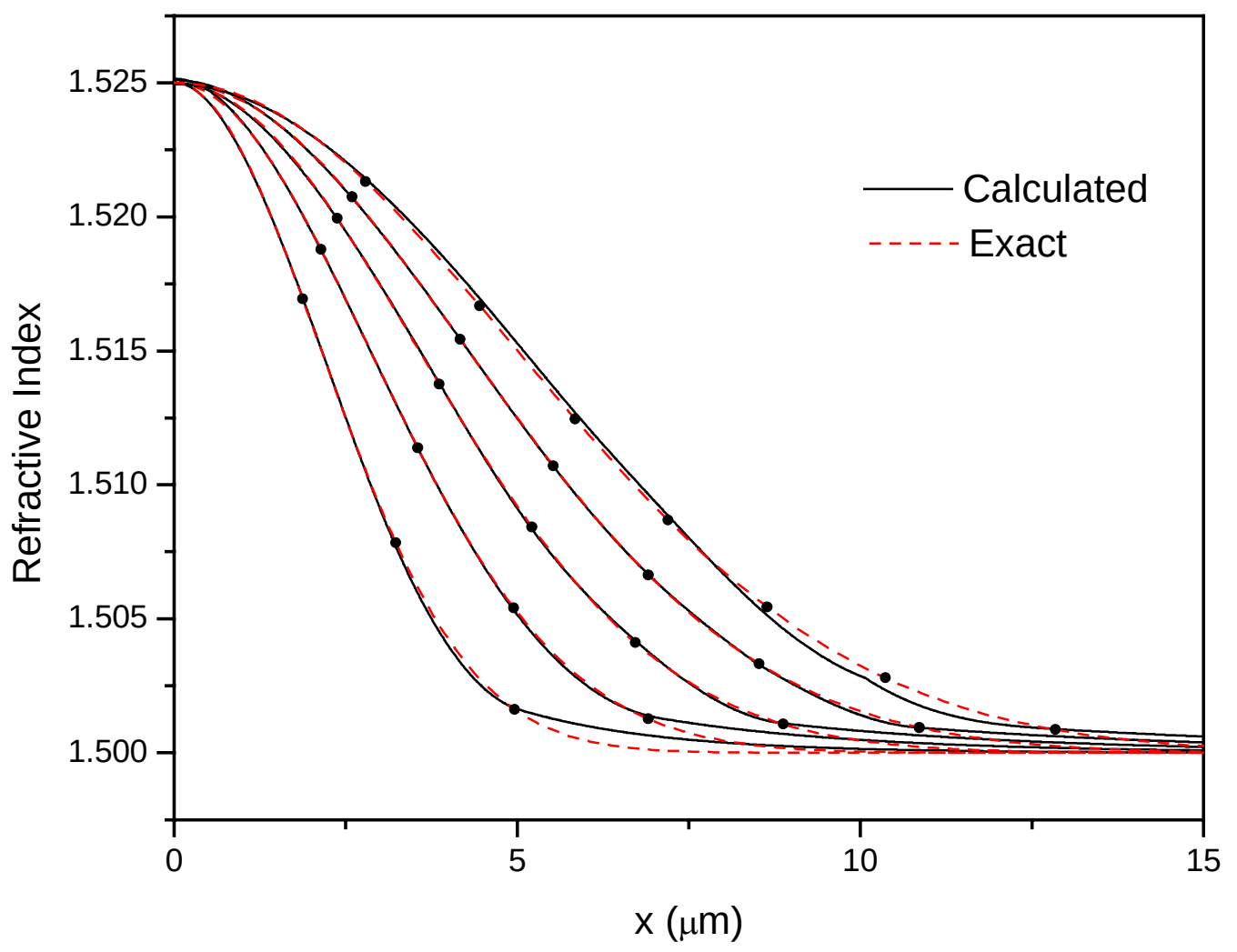

Fig.1(a) Recovery of Gaussian distribution by cubic spline function. $D_{gauss}$ varies from $7\mu m$ to $3\mu m$, allowing the waveguides to support 7~3 modes ( from the top down ).

Fig.1(b)

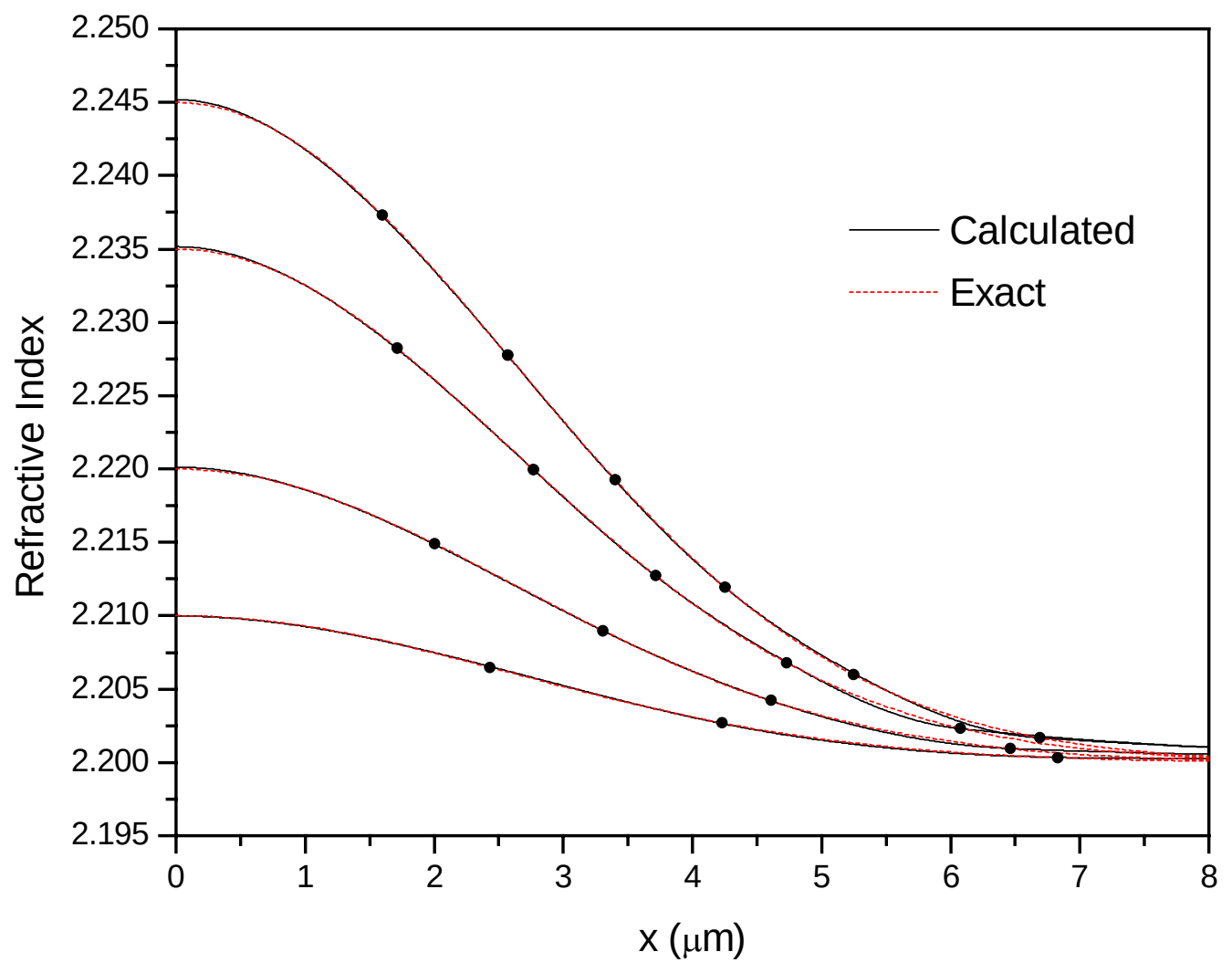

Fig.1(b) Recovery of error function distribution by cubic spline function. $n_{erf}$ = 0.045, 0.035, 0.02, 0.01 stimulate 6 modes, 5 modes, 4modes, 3 modes respectively (from the top down).

Fig.1(c)

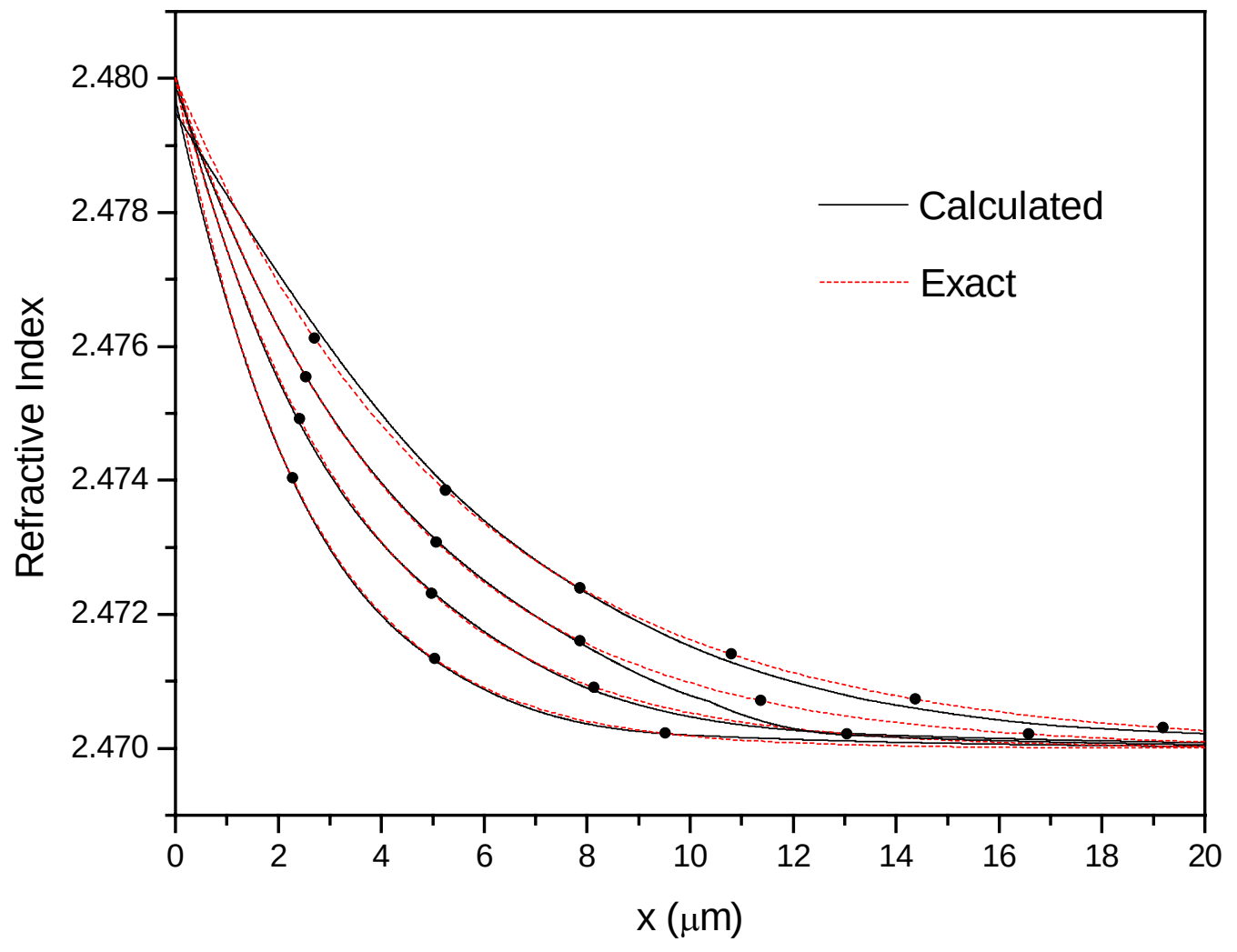

Fig.1(c) Recovery of exponential distribution by cubic spline function. The waveguides with $D_{\exp}$ of $5.5\mu m$, $4.3\mu m$, $3.4\mu m$, $2.5\mu m$ support 6,5,4,3 guiding modes respectively (from the top down).

Fig.1(d)

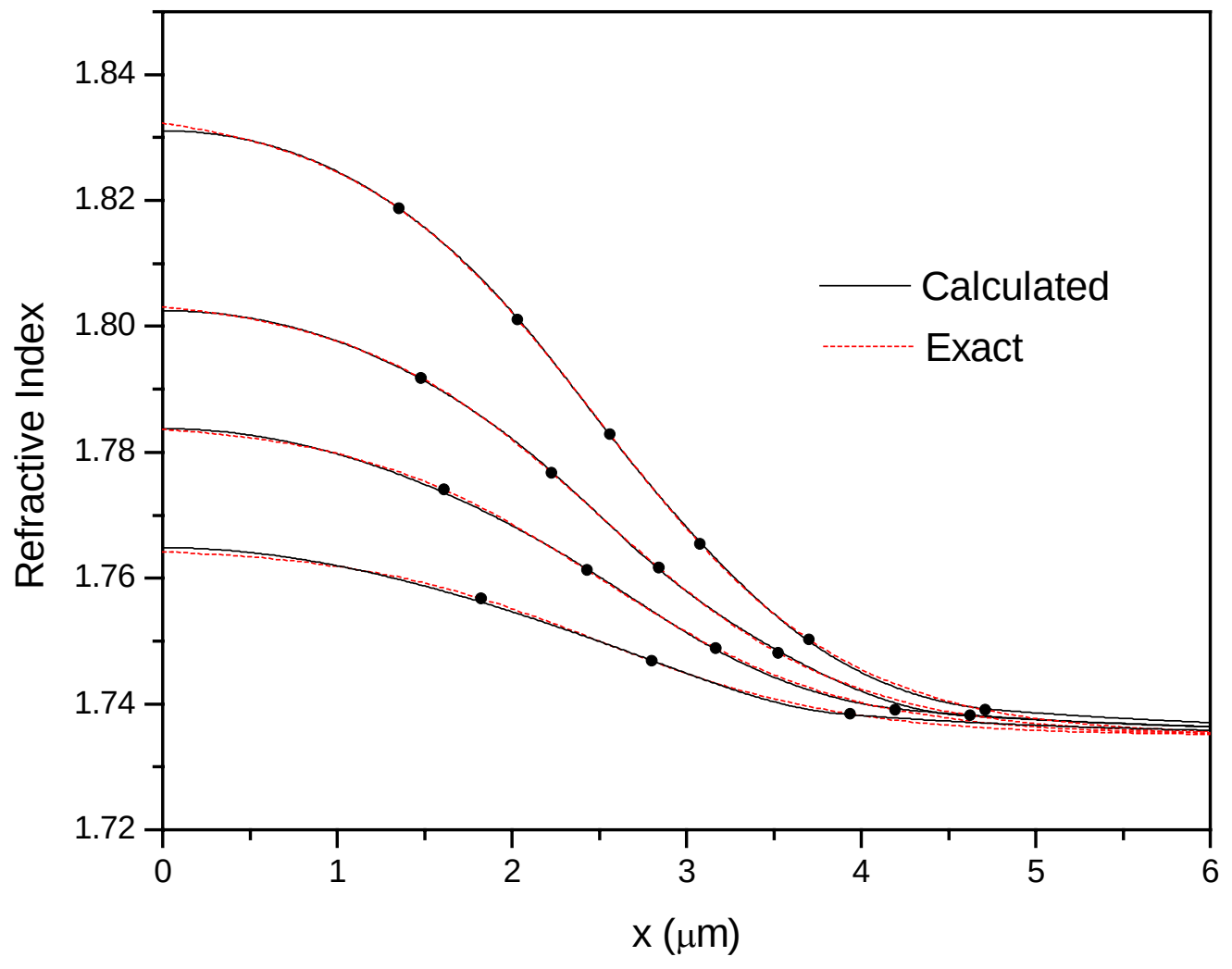

Fig.1(d) Recovery of slowly varying Fermi distribution by cubic spline function. $n_{\mathrm{fermi}}$ =0.1, 0.07, 0.05, 0.03 correspond with 6,5,4,3 stimulated guiding modes respectively (from the top down).

Fig.2

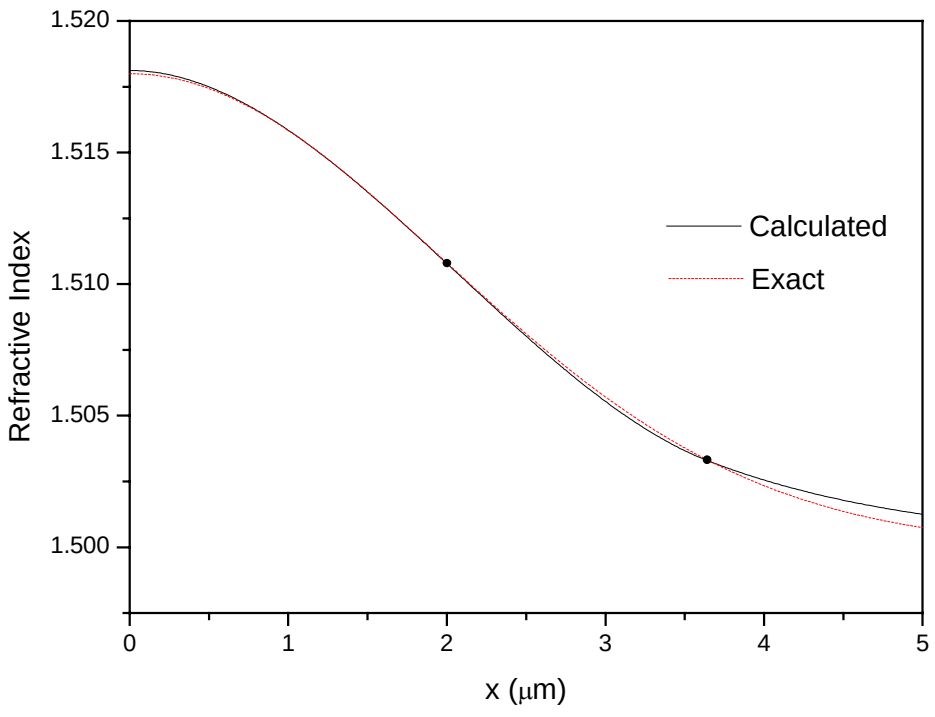

(a)

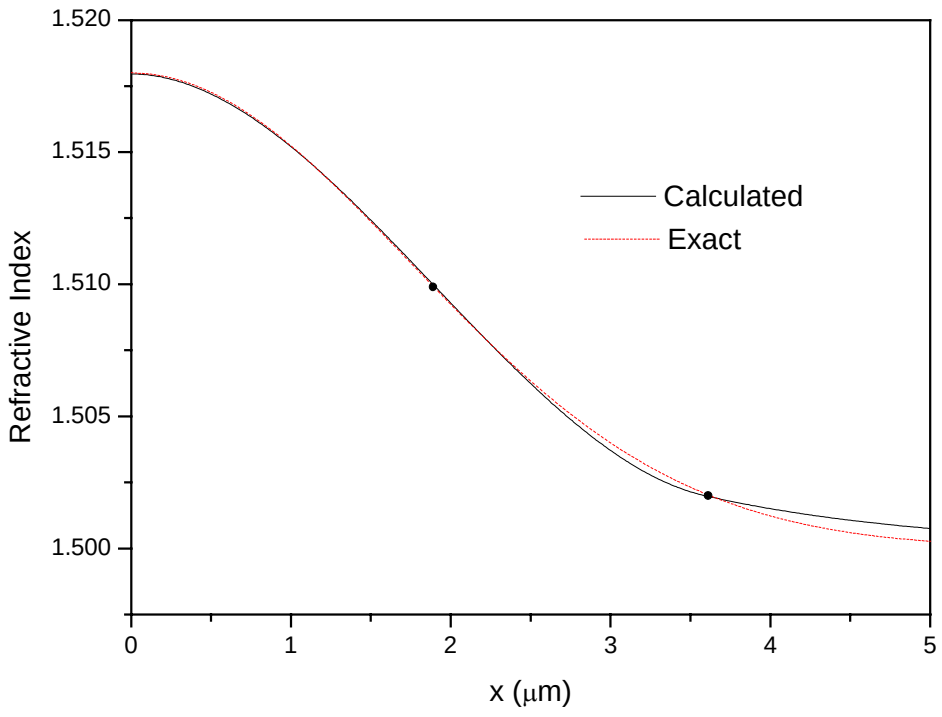

(b)

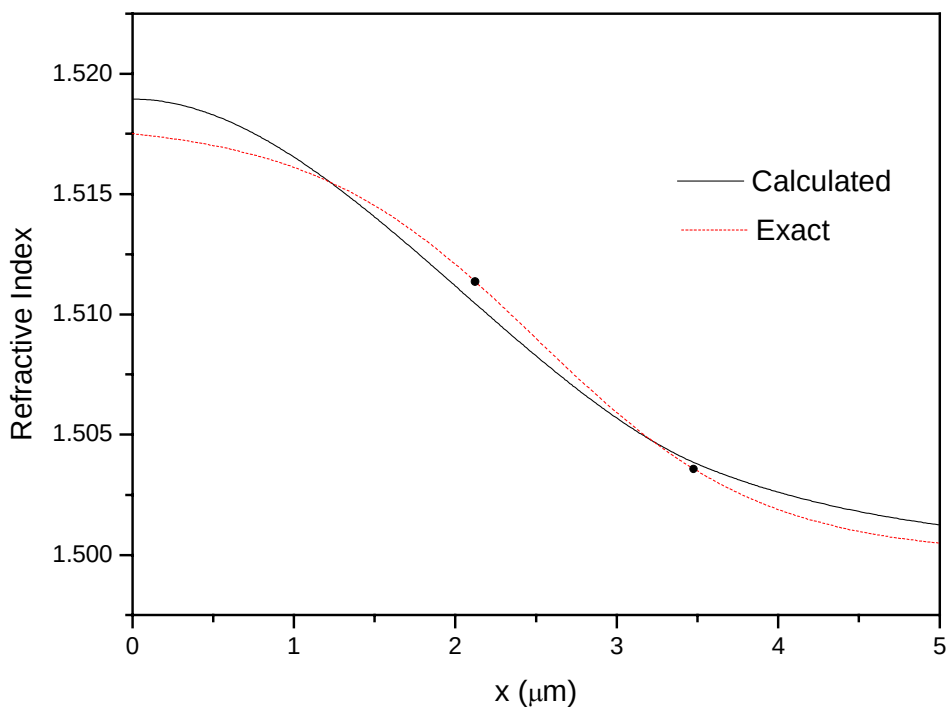

(c)

Fig. 2 Recovery of two-mode waveguide with (a) Gaussian profile, (b) error function profile, (c) slowly varying Fermi profile

Fig.3

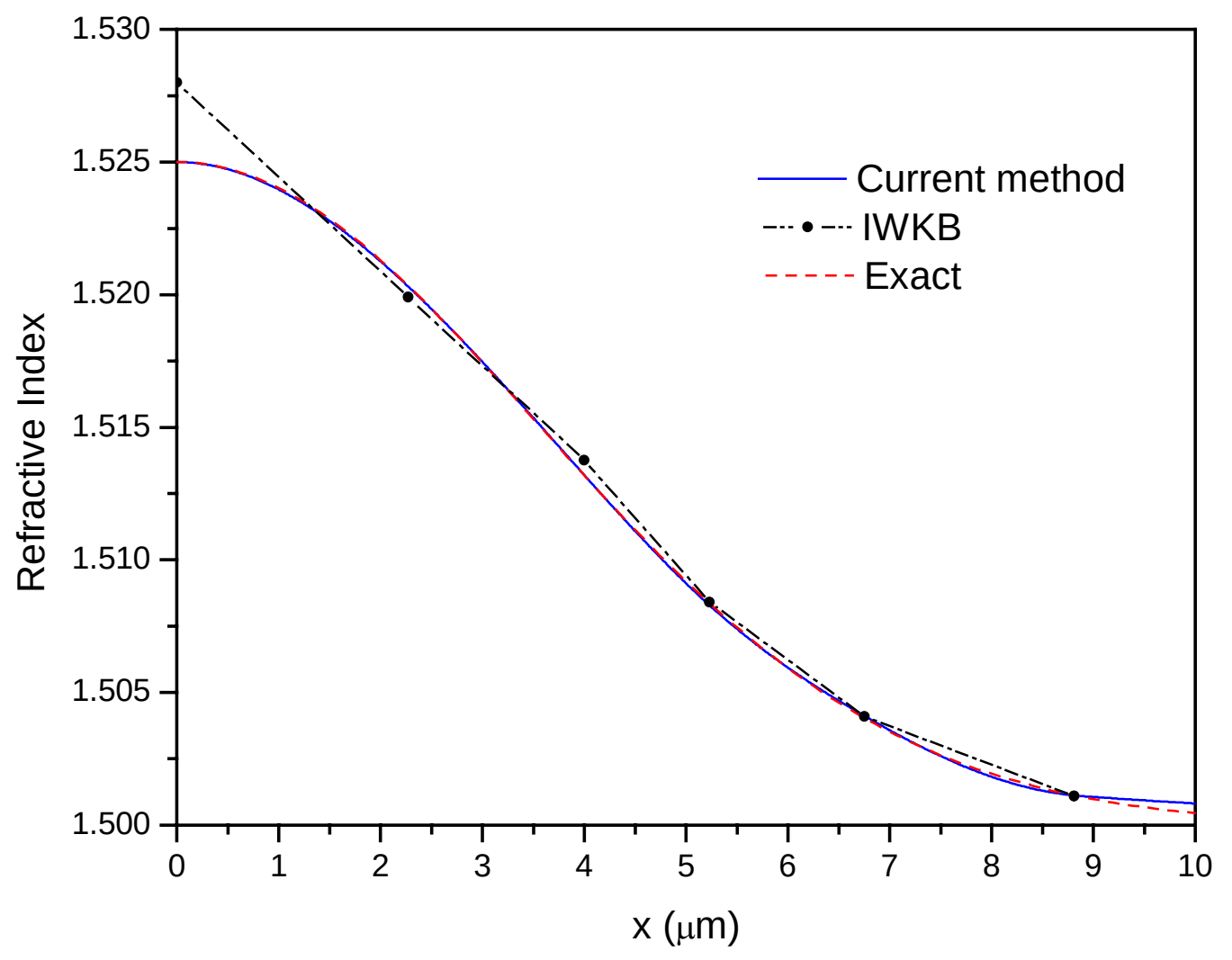

Fig.3. Comparison of recovered profiles from effective indexes by IWKB and the current method.